# Control of magnetization dynamics by spin Nernst torque

Arnab Bose, Ambika Shankar Shukla, Sutapa Dutta, Swapnil Bhuktare, Hanuman Singh, Ashwin A. Tulapurkar

Department of Electrical Engineering, Indian Institute of Technology-Bombay, Powai, Mumbai, India 400076

*Relativistic interaction between electron's spin and orbital angular momentum has provided efficient mechanism to control magnetization of nano-magnets. Extensive research has been done to understand and improve spin-orbit interaction driven torques generated by non-magnets while applying electric current. In this work, we show that heat current in non-magnet can also couple to its spin-orbit interaction to produce torque on adjacent ferromagnet. Hence, this work provides a platform to study spin-orbito-caloritronic effects in heavy metal/ferromagnet bi-layers.*

Since last few years considerable attention has been drawn to control the magnetization dynamics by pure spin current generated by spin Hall effect (SHE)[1,2,3,4] and interfacial magnetic fields[5,6] by Rashba effect. SHE and Rashba effects are relativistic phenomena which couple electron's spin and orbital motion and can be used to exert spin-orbit torques[7,8]. On the other hand, thermal gradient in ferromagnet can also create pure spin current[9,10,11,12,13] which can further produce thermal spin torques[14,15,16,17,18] and domain wall motion[19,20]. Conversion of heat current into spin current in a non-magnet has been shown recently via the spin Nernst effect (SNE)[21,22,23,24]. But an important question remains unanswered whether thermal gradient in non-magnet can generate spin toque owing to its spin-orbit coupling, which in turn could be used for manipulating magnetization. In this letter we demonstrate that, interplay of heat current and spin-orbit coupling in non-magnetic Platinum (Pt) can generate thermally driven spin-orbit torque (equivalent to spin Nernst torque (SNT)). SNT has been predicted recently[25,26] but it is lacking the experimental evidence. Here, we show that effective magnetic damping can be controlled by SNT while creating thermal gradient in Pt/$Ni_{81}Fe_{19}$ bilayer. This can open a new avenue to manipulate spins in magnetic nanostructures for technological applications[27,28,29].

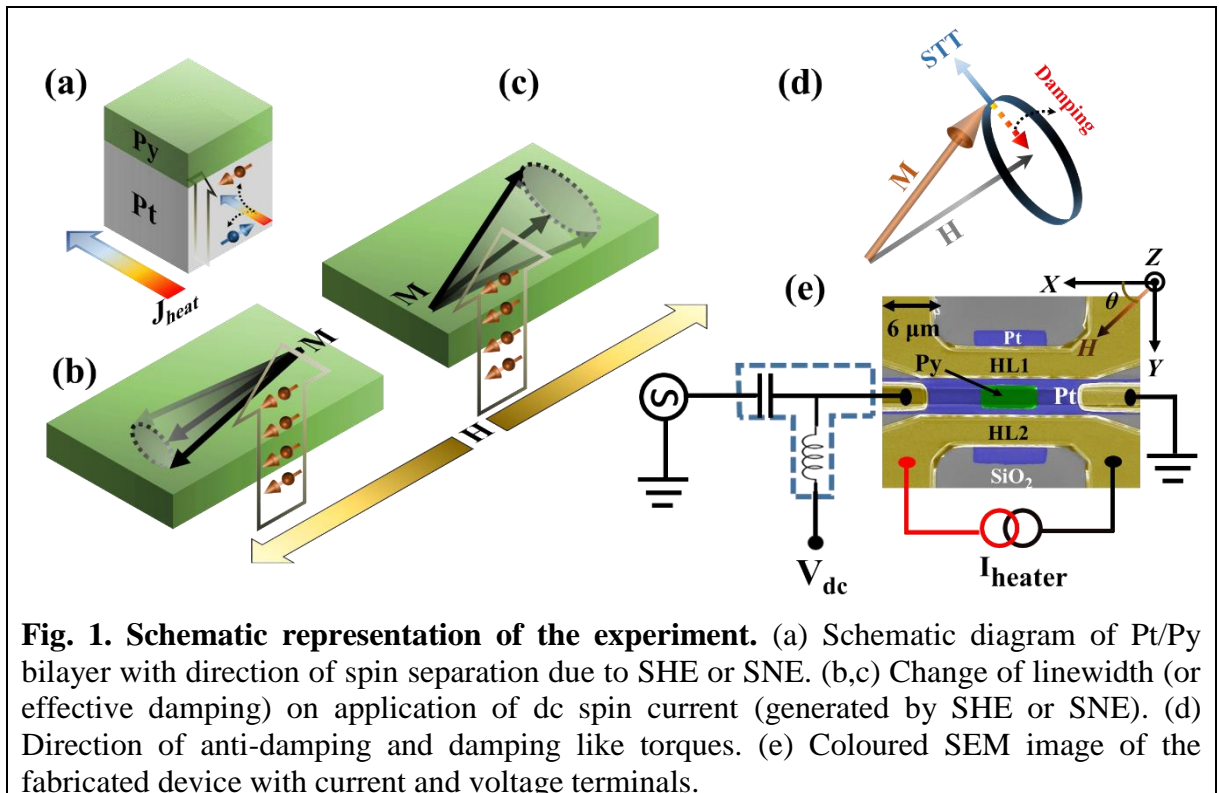

**Fig. 1. Schematic representation of the experiment.** (a) Schematic diagram of Pt/Py bilayer with direction of spin separation due to SHE or SNE. (b,c) Change of linewidth (or effective damping) on application of dc spin current (generated by SHE or SNE). (d) Direction of anti-damping and damping like torques. (e) Coloured SEM image of the fabricated device with current and voltage terminals.

While thermal gradient is established in heavy metal (Pt in this work), spins of opposite polarity separate out in a direction orthogonal to the direction of heat flow due to SNE as shown in Fig. 1(a). This situation is thermal analogous to SHE. Thus Pt converts heat current into pure spin current which is then injected into neighbouring ferromagnet (FM). If the injected spin current density is enough, spin torque is expected on the FM causing enhanced or reduced damping depending upon the direction of spin vectors absorbed by FM (Fig 1(b-c)). We compare the change in resonance linewidth of FM due to the spin torque generated by SNE and SHE by performing spin-torque ferromagnetic resonance (ST-FMR) experiment[2,4,6,30,31,32]. Basic working principle of ST-FMR is the following. Radio frequency (rf) current is applied along *X* axis and dc voltage is measured in the same terminals using a bias-T (AMR based detection of ST-FMR[2,32]) while external magnetic field is swept at angle $\theta$ with respect to *X*-axis (Fig. 1(e)). Pt converts rf charge current into rf spin current which is injected to ferromagnet ($Ni_{81}Fe_{19}$: Py here after). Radio frequency spin current and current induced rf fields excite the magnet to undergo small oscillation around its equilibrium position. Due to the AMR effect, resistance of the magnet (hence FM/HM stack) also oscillates. Homodyne mixture of RF applied current and RF resistance of the sample produces dc voltage. Advantage of ST-FMR is that at resonance large dc voltage can be obtained. This dc voltage is typically combination of symmetric Lorentzian ($V_S$) and anti-symmetric Lorentzian ($V_A$).

$$V_S = C_1 \frac{\Delta^2}{4(H-H_0)^2 + \Delta^2} \text{ and } V_A = C_2 \frac{4(H-H_0)\Delta}{4(H-H_0)^2 + \Delta^2}$$ where $C_1$ and $C_2$ are the amplitude of $V_S$

and $V_A$ respectively, $H$ is externally applied field, $H_0$ is resonant field position and $\Delta$ is the resonance linewidth (FWHM). $V_S$ indicates the contribution of spin current induced torque and $V_A$ indicates in-plane field induced torques. In Pt/Py bi-layers, the Oersted magnetic field is the dominant source of in-plane field.[2,32]. So charge to spin current conversion efficiency (or effective spin Hall angle) in HM can be quantified from $C_1/C_2$ ratio as following: $\theta'_{SH} = \frac{C_1}{C_2} \frac{e\mu_0 M_s t_{Pt} t_{Py}}{2\hbar} (1 + H_\perp / H_O)^{1/2}$ where $M_S$ is saturation magnetization, $t_{Pt}$, $t_{Py}$ are thickness of Pt and Py film respectively, $H_\perp$ is perpendicular magnetic anisotropy field. Now, if dc current is superimposed on the RF current then non-zero dc spin current is injected which can change the resonance linewidth of Py[2,3,32]. It also provides direct quantification of effective spin Hall angle as following:

$$\theta'_{SH} = \frac{\mu_0 M_S t_{Py}}{\cos\theta} \left(\frac{2e}{\hbar}\right) \left(\frac{\gamma}{4\pi f}\right) \left(H + \frac{1}{2}H_\perp\right) \left(\frac{d\Delta}{dJ_c}\right)_{Slope}$$ where $J_C$ is charge current density through Pt, $f$

is frequency of applied rf current, $\gamma$ is gyromagnetic ratio. In this work, we shave superimposed a dc heat current on rf charge current. We could modulate the resonance line width of Py which provides a direct evidence of control of magnetization dynamics by spin Nernst torque (SNT).

We fabricated the device as shown in Fig. 1(e). The crossbar is made of Pt (15 nm) and a rectangular shaped dot of Py (2 nm) is deposited at the centre of Pt crossbar. Top of Py was capped with Ta (1.5 nm). On the top and bottom lead of the Pt crossbar, two heater lines are fabricated which are electrically isolated by $SiO_x$ (30 nm) from Pt. Heater lines are prepared with Ta/Pt (60 nm). Numbers in bracket indicate thickness of metals. Entire fabrication is done by standard electron beam lithography, sputtering and lift-off technique. Before deposition of Py, surface of Pt was cleaned by Ar ion without breaking the vacuum. We have earlier shown that linewidth change can be sensitively measured in planar Hall structure doing ST-FMR [Ref. 32]. We follow the same approach in this work to compare the strength of spin Hall torque (SHT) and spin Nernst torque (SNT). Our detection method is the following. Radio frequency current is applied along *X*-axis; dc voltage is measured along same direction (hence AMR based detection) but dc heat current (or charge current) is passed along *Y*-axis to modulate the line width by SNT (or SHT) as shown in Fig. 1(e). Application of dc current (heat current or charge current) perpendicular to the direction of voltage measurement reduces noise and hence measurement sensitivity significantly increases as shown in Ref 32.

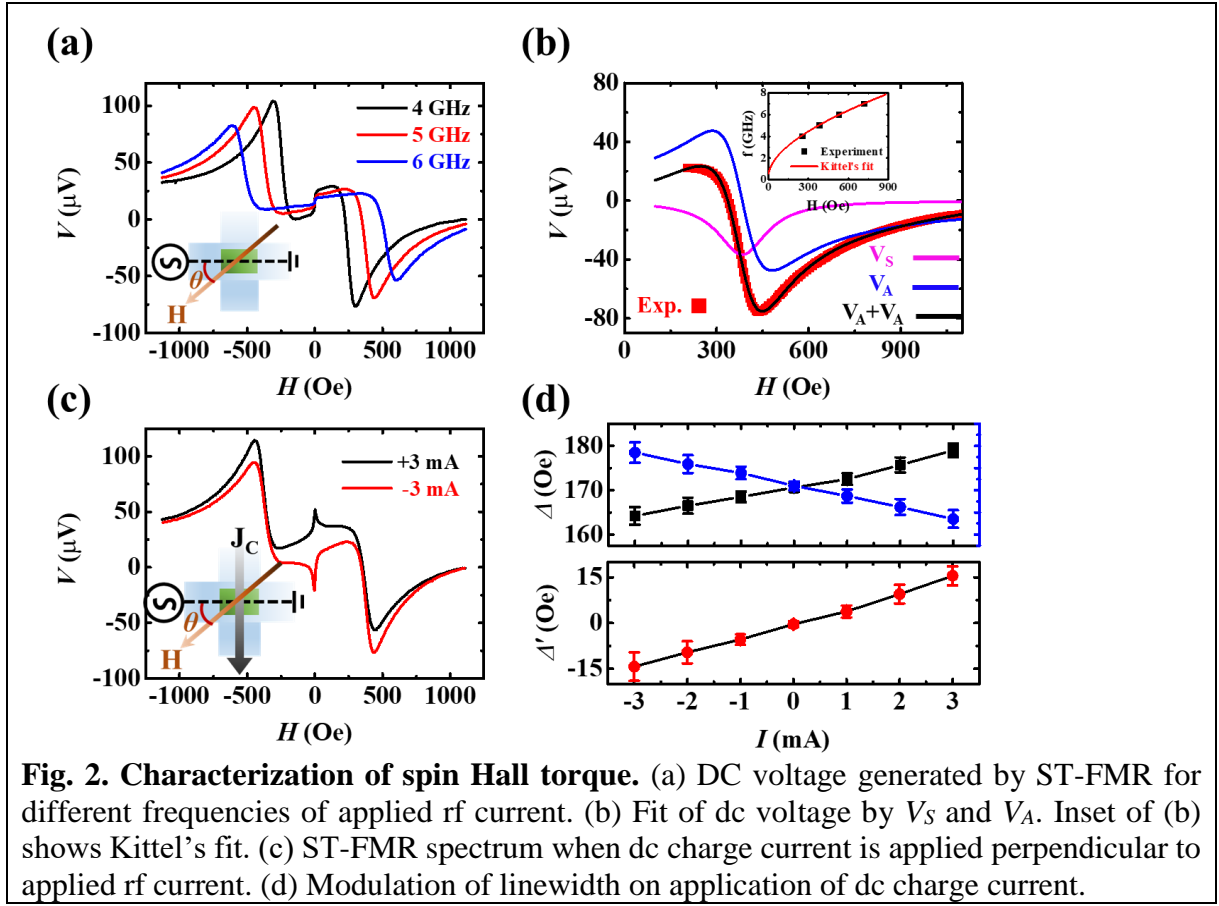

**Fig. 2. Characterization of spin Hall torque.** (a) DC voltage generated by ST-FMR for different frequencies of applied rf current. (b) Fit of dc voltage by $V_S$ and $V_A$. Inset of (b) shows Kittel's fit. (c) ST-FMR spectrum when dc charge current is applied perpendicular to applied rf current. (d) Modulation of linewidth on application of dc charge current.

Fig. 2 shows the characterization of spin-Hall torque by measuring ST-FMR. Fig. 2(a) shows the typical dc voltage spectrum as external magnetic field is swept for different frequencies of applied current. This dc voltage can be fitted to the sum of $V_S$ and $V_A$ (Fig. 2(b)). Red squares in Fig. 2(b) show the experimental data and black curve shows fitting which is sum of $V_S$ (pink curve) and $V_A$ (blue curve). The symmetric component confirms the spin-orbit torque generated by spin Hall effect. Inset of Fig. 2(b) shows the Kittel's fit for resonant magnetic fields and frequencies. From this we obtain $H_\perp = 8.05$ kOe (which corresponds to $M_s = 6.5 \times 10^5$ A/m.) Fig. 2(c) shows the voltage spectrum when dc charge current is applied orthogonal to the direction of rf current flow. We can clearly see the dominant change in the shape of the voltage signal as the linewidth significantly changes. For the positive current linewidth is more in positive field values and it is less in negative field values (vice versa for negative applied current). Our detection method is so sensitive that linewidth change is clearly visible in Fig. 2(c) itself. Linewidth ($\Delta$) as a function of applied dc current is shown in top panel of Fig. 2(d). It shows expected linear dependence as function of applied current. Bottom panel of Fig. 2(d) shows the difference in linewidth for $\theta=35^0$ and $\theta=215^0$ ($\Delta'=\Delta(35^0)-\Delta(215^0)$) as a function of applied current which also shows linear dependence. Bottom panel of Fig. 2(d) represents only the contribution of spin current induced damping change eliminating the overall heating effect if any. From this measurement of linewidth modulation, we can extract the effective spin Hall angle of Pt to be 0.12±0.06. Extracted value of spin Hall angle from $C_1/C_2$ ratio is somewhat lower. We have further noticed that in this planar Hall geometry when dc current is applied perpendicular to rf current modulation in $V_S$ is higher compared to the line-width modulation as a function of applied dc current. Same behaviour was also observed in our previous work (Fig. 3(c) of ref 32 and Fig. 2(c) in this work). However, we have verified that the line width modulation is the same irrespective of the detection methods (AMR or PHE based detection or hybrid planar Hall detection method). So in this work we quantify the strength of SHT and SNT in Pt by measuring linewidth change by injecting dc spin current by SHE and SNE respectively.

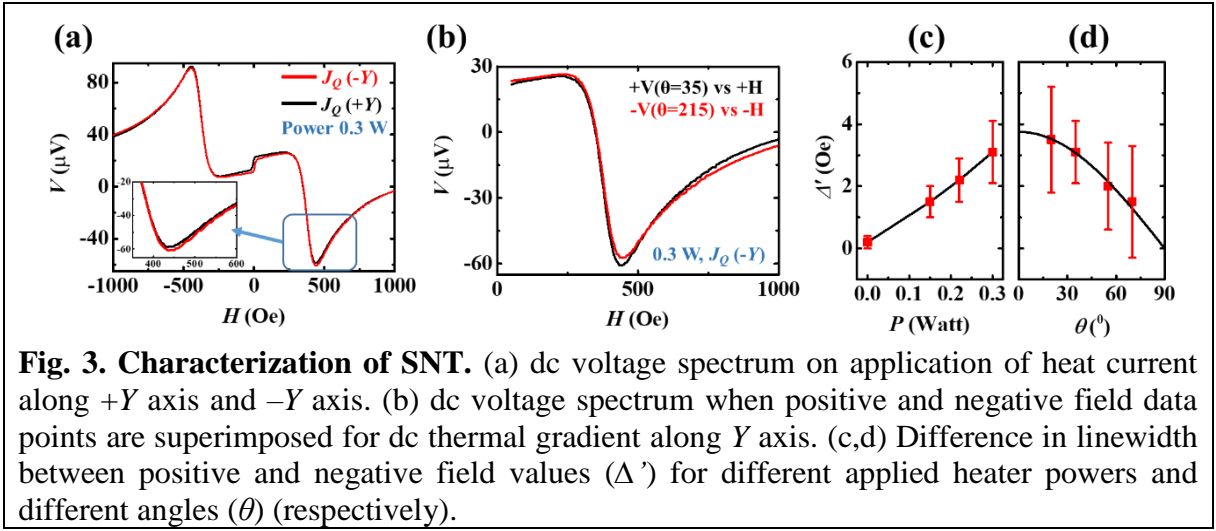

**Fig. 3. Characterization of SNT.** (a) dc voltage spectrum on application of heat current along +Y axis and –Y axis. (b) dc voltage spectrum when positive and negative field data points are superimposed for dc thermal gradient along Y axis. (c,d) Difference in linewidth between positive and negative field values ($\Delta'$) for different applied heater powers and different angles ($\theta$) (respectively).

Now we show the evidence of spin Nernst torque by passing dc heat current instead of applying dc charge current while doing ST-FMR as discussed above. Two different heater lines are fabricated on Hall bar (Fig. 1(e)) to create thermal gradient along ±Y-axis. When current flows in heaterline-1 (HL1) it becomes hot due to Joule's heating and most of the heat is carried by Pt below the heat line. This creates thermal gradient in Pt/Py bilayers along +Y axis. Similarly, when current is applied in heaterline-2 (HL2) thermal gradient is created along -Y axis. Fig. 3(a) shows the dc voltage spectrum generated by ST-FMR while thermal gradient is created along ±Y axis. We can see the difference in dc voltage spectrum for positive and negative thermal gradients. This cannot be explained by overall heating effect since overall heating would be same for both the direction of thermal gradients. All these measurements are performed when $\theta$ is $35^0$. In Fig. 3(b) we further show that when voltage signal of positive field and negative field is superimposed, there is distinct change in the shape of voltage spectrum (hence the linewidth) which further confirms the existence of SNT. Difference in the shape of voltage signal in Fig. 3(b) cannot be explained by overall heating effect as it would be same for both $35^0$ and $215^0$. As shown earlier (Fig. 2(d)), line-width difference for positive and negative field ($\Delta'=\Delta(\theta) - \Delta(\theta+180^0)$) is measured for different applied heater powers (Fig. 3(c)) and for different angles (Fig. 3(d)). $\Delta'$ is proportional to the heater power (Fig. 3(c)) and it closely follows $\cos\theta$ dependence (Fig. 3(d)). In our geometry (Fig. 1(a)) the line width modulation due to SNT is expected to be maximum at $\theta=0^0$, but ST-FMR signal becomes zero at $\theta=0^0$. We further confirmed that polarity of heater current has negligible effect on $\Delta'$ as heater line is electrically insulated from Pt hall bar and Py dot is quite far away (1.5 μm) from the heater line to get affected by magnetic (Oersted) field produced by the heater current. In our control experiment we have applied heater current in both HL1 and HL2 which would cause the same overall heating but fail to set up well directed thermal gradient. We found negligible effect on the line width in this case. Our observed results shown in Fig. 3 strongly supports the evidence of spin Nernst torque in Pt.

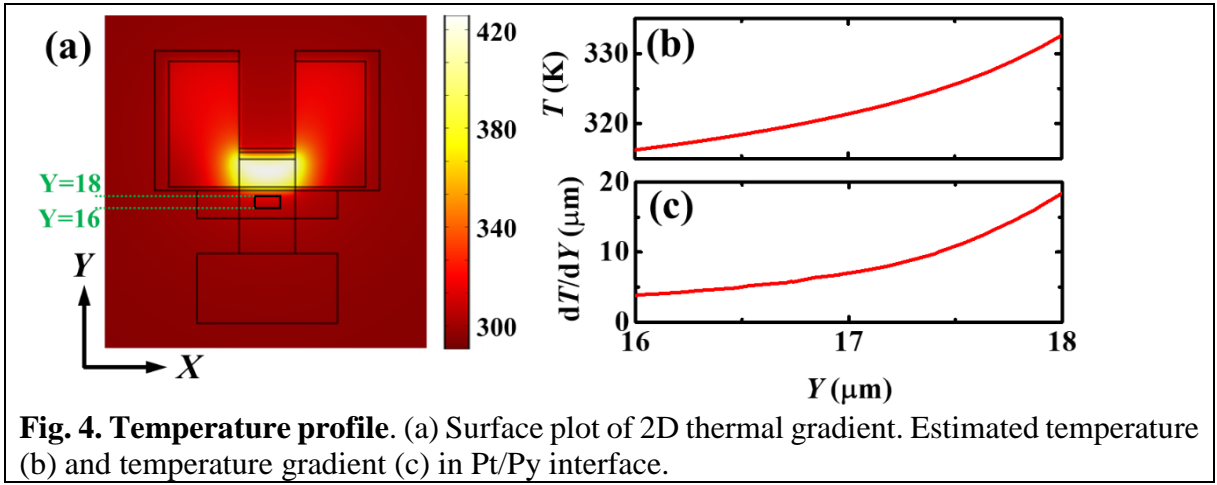

**Fig. 4. Temperature profile**. (a) Surface plot of 2D thermal gradient. Estimated temperature (b) and temperature gradient (c) in Pt/Py interface.

We have followed the same approach to find thermal gradient in Pt as reported in Ref 24. From the resistance value of the heater line, its overall temperature is known (on chip temperature calibration). Once the temperature of heater line is known temperature profile of Pt/Py interface can be calculated from COMSOL simulation (Fig. 4(b) and 4(c)). Overall temperature rise is around 25 K which is fairly small and hence we observe negligible contribution from overall heating effect. We consider the thermal gradient at the interface to be 15 K/μm while estimating SNT. Our estimated temperature gradient in this kind of geometry is in good agreement with previous results[20]. Comparing the line-width modulation by SHT and SNT we can quantify that 15 K/μm horizontal thermal gradient in Pt/Py interface is equivalent to the application of $4.9 \times 10^9$ A/m$^2$ amount of charge current density in Pt. This reported value of heat current to spin current conversion efficiency by SNE is consistent with our previous report of SNE[24] and comparable to reports by other groups[21,23].

In conclusion, we have demonstrated the control of magnetization dynamics by thermally driven spin Hall torque (or spin Nernst torque). We report approximately 0.9 % line-width change due to spin Nernst torque effect. It indicates that about 100 times more thermal gradient needs be created to achieve switching by spin Nernst torque which could be achieved in material having large spin Nernst angle and implementing efficient mechanism (such as Laser heating) to create large thermal gradient at nano scale. Further by using ferromagnets with less out-of-plane anisotropy, the thermal gradients required for switching can be reduced. These results will beneficial in energy harvesting sector where heat current in non-magnets can be utilized to write magnetic memories for non-volatile memory applications.

**References**

[1] J. Sinova, S. O. Valenzuela, J. Wunderlich, C. H. Back, and T. Jungwirth, "*Spin Hall effects*". Rev. Mod. Phys. 87, 1213 (2015)

[2] L. Q. Liu, T. Moriyama, D. C. Ralph, and R. A. Buhrman, "*Spin-Torque Ferromagnetic Resonance Induced by the Spin Hall Effect*", Phys. Rev. Lett. 106, 036601 (2011)

[3] K. Ando, S. Takahashi, K. Harii, K. Sasage, J. Ieda, S. Maekawa, and E. Saitoh, "*Electric Manipulation of Spin Relaxation Using the Spin Hall Effect*", Phys. Rev. Lett. 101, 036601 (2008)

[4] L. Liu, Chi-Feng Pai, D. C. Ralph, and R. A. Buhrman, Magnetic Oscillations Driven by the Spin Hall Effect in 3-Terminal Magnetic Tunnel Junction Devices, Phys. Rev. Lett. **109**, 186602 (2012).

[5] I. M. Miron, G. Gaudin, S. Auffret, B. Rodmacq, A. Schuhl, S. Pizzini, J. Vogel, P. Gambardella. "*Current-driven spin torque induced by the Rashba effect in a ferromagnetic metal layer*", Nat. Mater. 9, 230 (2010)

[6] A. Bose, H. Singh, V. K. Kushwaha, S. Bhuktare, S. Dutta, and A. A. Tulapurkar. "*Sign Reversal of Field-like Spin-Orbit Torque in an Ultrathin Cr/Ni Bilayer*", Phys. Rev. Appl. 9, 014022 (2018)

[7] Anjan Soumyanarayanan, Nicolas Reyren, Albert Fert, and Christos Panagopoulos, "*Emergent phenomena induced by spin-orbit coupling at surfaces, and interfaces*", Nature (London) **539**, 509 (2016)

[8] A. Manchon, H. C. Koo, J. Nitta, S. M. Frolov, and R. A. Duine, "*New perspectives for Rashba spin orbit coupling*", Nat. Mater. **14**, 871 (2015)

[9] K. Uchida, H. Adachi, T. Ota, H. Nakayama, S. Maekawa, and E. Saitoh. "*Observation of longitudinal spin-Seebeck effect in magnetic insulators*", Appl. Phys. Lett. **97**, 172505 (2010)

[10] G. E. W. Bauer, E. Saitoh and B. J. Van Wees. "*Spin Caloritronics*", Nat. Mater. **11**, 391 (2012)

[11] M. Walter, J. Walowski, V. Zbarsky, M. M€unzenberg, M. Sch€afers, D. Ebke, G. Reiss, A. Thomas, P. Peretzki, M. Seibt et al., "*Seebeck effect in magnetic tunnel junctions*," Nat. Mater. **10**, 742–746 (2011)

[12] S. Jain, D. D. Lam, A. Bose, H. Sharma, V. R. Palkar, C. V. Tomy, Y.Suzuki, and A. A. Tulapurkar, "*Magneto-Seebeck effect in spin-valve with heat current in-plane thermal gradient*", AIP Adv. **4**, 127145 (2014)

[13] M. Schmid, S. Srichandan, D. Meier, T. Kuschel, J.-M. Schmalhorst, M. Vogel, G. Reiss, C. Strunk and C. H. Back. "*Transverse Spin Seebeck Effect versus Anomalous and Planar Nernst Effects in Permalloy Thin Films*", Phys. Rev. Lett. **111**, 187201 (2013)

[14] H. Yu, S. Granville, D. P. Yu, and Ph. J. Ansermet, "*Evidence for thermal spin-transfer torque*", Phys. Rev. Lett. **104**, 146601 (2010)

[15] M. G. Choi, H. C. Moon, C. B. Min, J. K. Lee, and G. D. Cahill, "*Thermal spin-transfer torque driven by the spin-dependent Seebeck effect in metallic spin-valves*", Nat. Phys. **11**, 576–582 (2015).

[16] A. Pushp, T. Phung, C. Rettner, B. P. Hughes, S. H. Yang, and S. S. P. Parkin, "*Giant thermal spin torque-assisted magnetic tunnel junction switching*," Proc. Natl. Acad. Sci. U.S.A. **112**, 6585–6590 (2015).

[17] A. Bose et al., "*Observation of thermally driven field-like spin torque in magnetic tunnel junctions*", Appl. Phys. Lett. **109**, 032406 (2016)

[18] M. B. Jungfleisch, T. An, K. Ando, Y. Kajiwara, K. Uchida, V. I. Vasyuchka, A. V. Chumak, A. A. Serga, E. Saitoh and B. Hillebrands. "*Heat-induced damping modification in yttrium iron garnet/platinum heterostructures*", Appl. Phys. Lett. **102**, 062417 (2013)

[19] W. Jiang, P. Upadhyaya, Y. Fan, J. Zhao, M. Wang, L. Chang, M. Lang, K. L. Wong, M. Lewis, Y.-T. Lin et. al. "*Direct Imaging of Thermally Driven Domain Wall Motion in Magnetic Insulators*", Phys. Rev. Lett. **110**, 177202 (2013)

[20] P. Krzysteczko, X. Hu, N. Liebing, S. Sievers, and H. W. Schumacher. "*Domain wall magneto-Seebeck effect*", Phys. Rev. B **92**, 140405(R) (2015)

[21] S. Meyer, Y.-T. Chen, S. Wimmer, M. Althammer, T. Wimmer, R. Schlitz, S. Geprägs, H. Huebl, D. Ködderitzsch, H. Ebert, G.E.W. Bauer, R. Gross, and S.T.B. Goennenwein, "*Spin Nernst effect*", Nat. Mater. **16**, 977 (2017)

[22] P. Sheng, Y. Sakuraba, Y. Lau, S. Takahashi, S. Mitani, and M. Hayashi, "*The spin Nernst effect in tungsten*", Sci. Adv. **3**, e1701503 (2017)

[23] D.J. Kim, C.Y. Jeon, J. G. Choi, J.W. Lee, S. Surabhi, J.R. Jeong, K.J. Lee, and B.G. Park, Nat. "*Observation of transverse spin Nernst magnetoresistance induced by thermal spin current in ferromagnet/non-magnet bilayers*", Nat. Commun. **8**, 1 (2017)

[24] A. Bose, S. Bhuktare, H. Singh, S. Dutta, V. Achanta, A. A. Tulapurkar, "*Direct detection of spin Nernst effect in Pt*", Appl. Phys. Lett. **112**, 162401 (2018)

[25] G. Geranton, F. Freimuth, S. Blugel, and Y. Mokrousov. "*Spin-orbit torques in L10-FePt/Pt thin films driven by electrical and thermal currents*", Phys. Rev. B **91**, 014417 (2015)

[26] A. A. Kovalev and V. Zyuzin. "*Spin torque and Nernst effects in Dzyaloshinskii-Moriya ferromagnets*", Phys. Rev. B **93**, 161106(R) (2016).

[27] S. Bhuktare, H. Singh, A. Bose, and A. A. Tulapurkar, "*Spintronic Oscillator Based on Spin-Current Feedback Using the Spin Hall Effect*", Phys. Rev. Appl. **7**, 014022 (2017)

[28] Swapnil Bhuktare, Arnab Bose, Hanuman Singh & Ashwin A. Tulapurkar, "*Gyrator Based on Magnetoelastic Coupling at a Ferromagnetic/Piezoelectric Interface*", Scientific Reports | 7: 840 | DOI:10.1038/s41598-017-00960-9

[29] H. Singh et. al. "*Integer, Fractional, and Sideband Injection Locking of a Spintronic Feedback Nano-Oscillator to a Microwave Signal*", Phys. Rev. Appl. **8**, 064011 (2017)

[30] A. A. Tulapurkar, Y. Suzuki, A. Fukushima, H. Kubota, H. Maehara, K. Tsunekawa, D. D. Djayaprawira, N. Watanabe, and S. Yuasa, "*Spin Torque Diode effect in Magnetic Tunnel Junctions*", Nature (London) **438**, 339 (2005)

[31] A. Bose, S. Dutta, S. Bhuktare, H. Singh, and A. A. Tulapurkar, "*Sensitive measurement of spin orbit torque driven ferromagnetic resonance detected by planar Hall geometry*", Appl. Phys. Lett. 111, 162405 (2017)

[32] A. Bose, D. D. Lam, S. Bhuktare, H. Singh, S. Miwa and A. Tulapurkar, "*Observation of anomalous spin-torque generated by a ferromagnet*", arXiv:1706.07245 (2017)

# Supplementary information

Arnab Bose, Ambika Shukla, Sutapa Dutta, Swapnil Bhuktare, Hanuman Singh, Ashwin A Tulapurkar

*Department of Electrical Engineering, Indian Institute of Technology-Bombay, Powai, Mumbai, India 400076*

**S1. Temperature calibration**

Thermal gradient is obtained by doing on-chip calibration. As current is passed through heater line it becomes hot due to Joule's heating and its resistance increases. Hence resistance of heater line provides information of average temperature in heater line. Same approach was adopted in our previous work [S1] and by other researchers [S2,S3,S4]. Heater line was made of Ta/Pt (~50 nm). Its resistance changes from 50 Ω to 65 Ω on application of maximum heater power (0.3 W). This corresponds to overall temperature of heater line to be approximately 420 K which is obtained from the temperature dependent resistance measurement of the heater line. We observed minor change in resistance in Pt Hall bar which contains rectangular Py dot at the centre. It indicates that the temperature of horizontal Pt line does not increase much and the heat is locally centred near the heater line. To get exact temperature profile, COMSOL simulation is done with heat transfer module with following boundary conditions: (1) temperature of heater line is 420 K (experimentally obtained) and (2) temperature of the bottom of Si is 293 (lab temperature).

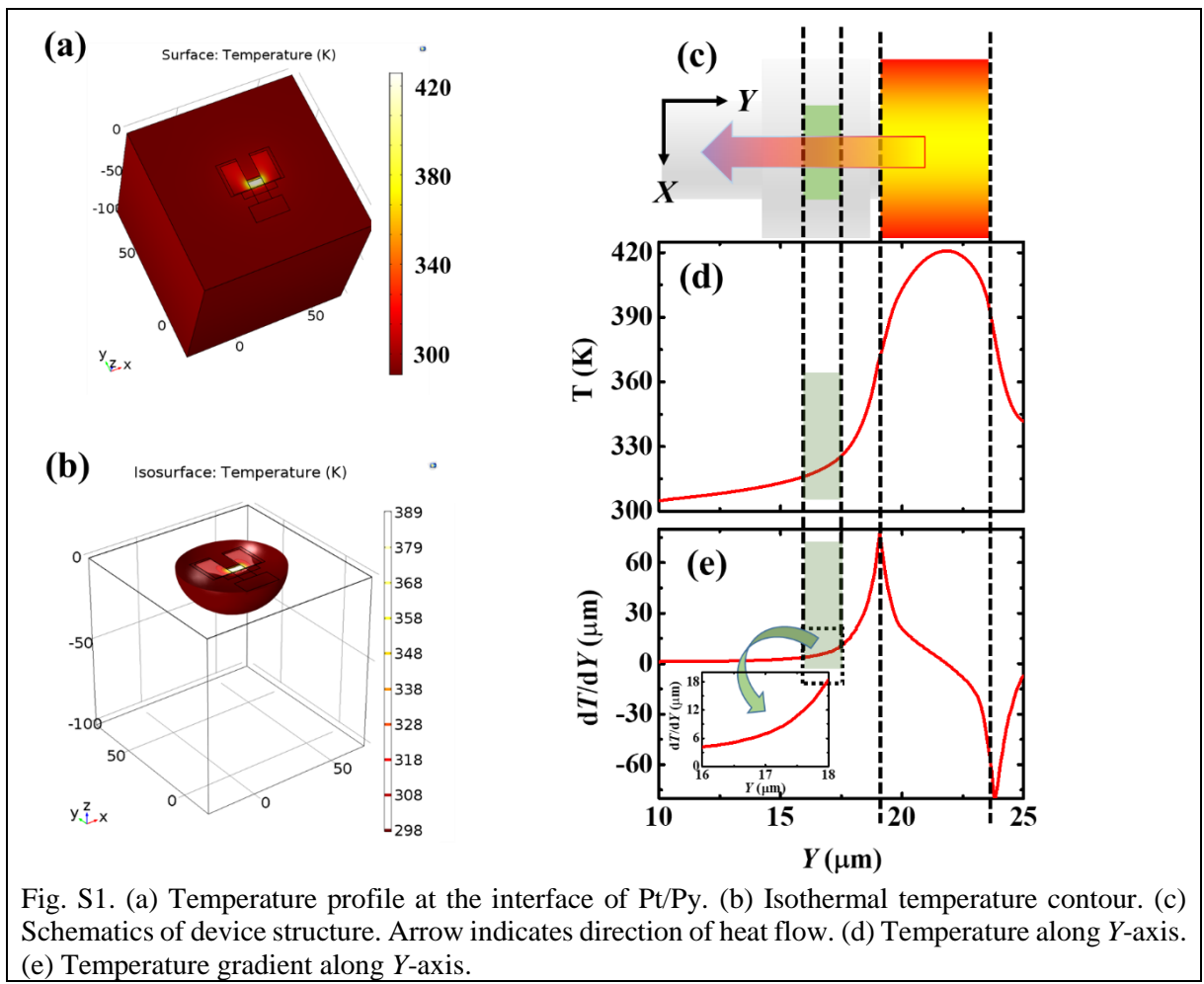

Fig. S1. (a) Temperature profile at the interface of Pt/Py. (b) Isothermal temperature contour. (c) Schematics of device structure. Arrow indicates direction of heat flow. (d) Temperature along *Y*-axis. (e) Temperature gradient along *Y*-axis.

Fig. S1(a) shows the surface temperature. Bright yellow colour indicates hot region and temperature of heater is set 420 K. Deep red indicates the colder region. Fig. S2(b) shows the isothermal contour which shows that heat is also localized in very small region (approx. 10×10×10 μm$^3$) as expected. Fig. S1(d) shows the temperature as along Y axis which shows exponential decay of thermal gradient (Fig. S1(e)). Temperature becomes around 300 K when we go 10 μm away from the heater line. It shows that overall temperature of Py is slightly more than room temperature (~325 K). Maximum thermal gradient along Y axis in Pt/Py interface is nearly 20 K/μm. Estimated temperature profile is in good agreement with previous works [S1-S6]. We have considered below parameters in simulation. We have also checked that slight variation of these parameter does not influence the simulated result much. Finally estimated thermal gradient will be in order tens of K/μm in this kind of geometry.

| Material | Pt | Si | SiO2 |
| --- | --- | --- | --- |
| Thermal conductivity (SI unit) | 85 | 120 | 1.4 |

References

[S1] A. Bose, S. Bhuktare, H. Singh, S. Dutta, V. Achanta, A. A. Tulapurkar. Appl. Phys. Lett. (accepted)

[S2] P. Krzysteczko, X. Hu, N. Liebing, S. Sievers, and H. W. Schumacher. Domain wall magneto-Seebeck effect. Phys. Rev. B 92, 140405(R) (2015)

[S3] S. Meyer, Y.-T. Chen, S. Wimmer, M. Althammer, T. Wimmer, R. Schlitz, S. Geprägs, H. Huebl, D. Ködderitzsch, H. Ebert, G.E.W. Bauer, R. Gross, and S.T.B. Goennenwein, Nat. Mater. **16**, 977 (2017)

[S4] A. Pushp, T. Phung, C. Rettner, B. P. Hughes, S. H. Yang, and S. S. P. Parkin, "Giant thermal spin torque-assisted magnetic tunnel junction switching," Proc. Natl. Acad. Sci. U.S.A. 112, 6585–6590 (2015).

[S5] A. Slachter, L. F. Bekker, J.-P. Adam, and J. B. van Wees, "Thermally driven spin injection from a ferromagnet into a non-magnetic metal," Nat. Phys. 6, 879–882 (2010)

[S6] M. Walter, J. Walowski, V. Zbarsky, M. M€ unzenberg, M. Sch€afers, D. Ebke, G. Reiss, A. Thomas, P. Peretzki, M. Seibt et al., "Seebeck effect in magnetic tunnel junctions," Nat. Mater. 10, 742–746 (2011)